\documentclass[%
 reprint,
 superscriptaddress,
 amsmath,amssymb,
 aps,
 prl,
]{revtex4-2}

\usepackage{epstopdf}
\usepackage[utf8]{inputenc}

\usepackage{graphicx}
\usepackage{dcolumn}
\usepackage{bm}
\usepackage{booktabs}
\usepackage{subfigure}
\usepackage{placeins}
\usepackage[USenglish]{babel}
\usepackage{hyphenat}
\newcommand{\sn}{$S_n$}

\newcommand{\qbeta}{$Q_{\beta}$}
\newcommand{\ponen}{$P_{1n}$}
\newcommand{\pxn}{$P_{xn}$}
\newcommand{\ptwon}{$P_{2n}$}

\newcommand{\halflife}{$T_{1/2}$}
\newcommand*{\myisoSimp}[2]{$^{#1}\mathrm{#2}$}
\newcommand{\rprocess}{\mbox{$r$~process}}
\newcommand{\sprocess}{\mbox{$s$~process}}

\usepackage{longtable}
\newcolumntype{L}{>{$}l<{$}} 

\usepackage[breaklinks]{hyperref}
\usepackage[hyphenbreaks=none]{breakurl}

\begin{document}

\title{Impact of newly measured $\beta$\nobreakdash-delayed neutron emitters around \myisoSimp{78}{Ni} on light element nucleosynthesis in the neutrino-wind following a neutron star merger }

\author{ A. Tolosa-Delgado  } 
    \email[Corresponding author: ]{alvaro.tolosa.delgado@cern.ch}
    \affiliation{Instituto de Fisica Corpuscular (CSIC-Universitat de Valencia), E-46980 Paterna, Spain}
    \affiliation{Accelerator Laboratory, Department of Physics, University of Jyväskylä, FIN-40014 Jyväskylä, Finland}
    \affiliation{European Organization for Nuclear Research (CERN) Switzerland}
\author{J. L. Tain } 
    \affiliation{Instituto de Fisica Corpuscular (CSIC-Universitat de Valencia), E-46980 Paterna, Spain}
\author{M. Reichert}
    \affiliation{Departament d'Astronomia i Astrof\'isica, Universitat de Val\`encia, E-46100 Burjassot, Spain}
\author{A. Arcones } 
    \affiliation{Institut für Kernphysik, Technische Universität Darmstadt, D-64289 Darmstadt, Germany}   
    \affiliation{GSI Helmholtzzentrum für Schwerionenforschung GmbH,  D-64291 Darmstadt, Germany}  
    \affiliation{Max-Planck-Institut für Kernphysik, Saupfercheckweg 1, 69117 Heidelberg, Germany}
\author{M. Eichler}
    \affiliation{Institut für Kernphysik, Technische Universität Darmstadt, D-64289 Darmstadt, Germany}   
\author{B. C. Rasco } 
    \affiliation{Physics Division, Oak Ridge National Laboratory, Oak Ridge, TN 37831, USA} 
\author{N. T. Brewer } 
    \affiliation{Physics Division, Oak Ridge National Laboratory, Oak Ridge, TN 37831, USA}    
\author{K.P. Rykaczewski } 
    \affiliation{Physics Division, Oak Ridge National Laboratory, Oak Ridge, TN 37831, USA}  
\author{R. Yokoyama } 
    \affiliation{Department of Physics \& Astronomy, University of Tennessee , Knoxville, TN 37996-1200 USA}  
\author{R. Grzywacz } 
    \affiliation{Department of Physics \& Astronomy, University of Tennessee , Knoxville, TN 37996-1200 USA}
\author{I. Dillmann } 
    \affiliation{TRIUMF, 4004 Wesbrook Mall, Vancouver, BC, V6T 2A3, Canada}  
    \affiliation{Department of Physics and Astronomy, University of Victoria, Victoria, BC,  V8P 5C2, Canada}  
\author{ J. Agramunt } 
    \affiliation{Instituto de Fisica Corpuscular (CSIC-Universitat de Valencia), E-46980 Paterna, Spain}  
\author{D. S. Ahn } 
    \affiliation{RIKEN Nishina Center, Wako, Saitama 351-0198, Japan}  
    \affiliation{Center for Exotic Nuclear Studies, Institute for Basic Science, Daejeon
34126, Republic of Korea}
\author{A. Algora } 
    \affiliation{Instituto de Fisica Corpuscular (CSIC-Universitat de Valencia), E-46980 Paterna, Spain}   
    \affiliation{HUN-REN Institute for Nuclear Research, Debrecen Pf. 51, H-4001, Hungary}  
 \author{H. Baba } 
    \affiliation{RIKEN Nishina Center, Wako, Saitama 351-0198, Japan}  
 \author{S. Bae } 
    \affiliation{Department of Physics \& Astronomy, Seoul National University, Seoul, 08826, Republic of Korea}  
    \affiliation{Center for Nuclear Study, The University of Tokyo, Hirosawa 2-1, Wako, 351-0198, Saitama, Japan}
 \author{C. G. Bruno} 
    \affiliation{School of Physics and Astronomy, The University of Edinburgh, Edinburgh, EH9 3FD, United Kingdom}  
 \author{R. Caballero Folch } 
    \affiliation{TRIUMF, 4004 Wesbrook Mall, Vancouver, BC, V6T 2A3, Canada} 
 \author{F. Calvino } 
    \affiliation{Universitat Politecnica de Catalunya (UPC), Barcelona, Spain}  
 \author{P. J. Coleman-Smith } 
    \affiliation{STFC Daresbury Laboratory, Daresbury, Warrington WA4 4AD, United Kingdom}  
 \author{G. Cortes } 
    \affiliation{Universitat Politecnica de Catalunya (UPC), Barcelona, Spain}  
 \author{T. Davinson } 
    \affiliation{School of Physics and Astronomy, The University of Edinburgh, Edinburgh, EH9 3FD, United Kingdom}  
 \author{C. Domingo-Pardo } 
    \affiliation{Instituto de Fisica Corpuscular (CSIC-Universitat de Valencia), E-46980 Paterna, Spain}  
 \author{A. Estrade } 
    \affiliation{Central Michigan University, Mount Pleasant, MI 48859, USA}  
 \author{N. Fukuda } 
    \affiliation{RIKEN Nishina Center, Wako, Saitama 351-0198, Japan}  
 \author{S. Go } 
    \affiliation{Department of Physics \& Astronomy, University of Tennessee , Knoxville, TN 37996-1200 USA}  
    \affiliation{RIKEN Nishina Center, Wako, Saitama 351-0198, Japan} 
 \author{C. J. Griffin } 
    \affiliation{School of Physics and Astronomy, The University of Edinburgh, Edinburgh, EH9 3FD, United Kingdom}  
 \author{J. Ha } 
    \affiliation{Department of Physics \& Astronomy, Seoul National University, Seoul, 08826, Republic of Korea}  
 \author{O. Hall } 
    \affiliation{School of Physics and Astronomy, The University of Edinburgh, Edinburgh, EH9 3FD, United Kingdom}  
 \author{L. Harkness-Brennan } 
    \affiliation{Department of Physics, University of Liverpool, Liverpool, L69 7ZE, United Kingdom}  
 \author{T. Isobe } 
    \affiliation{RIKEN Nishina Center, Wako, Saitama 351-0198, Japan}  
 \author{D. Kahl } 
    \affiliation{School of Physics and Astronomy, The University of Edinburgh, Edinburgh, EH9 3FD, United Kingdom}  
    \affiliation{Extreme Light Infrastructure-Nuclear Physics (ELI-NP)/Horia Hulubei National Institute
for Physics and Nuclear Engineering (IFIN-HH), Str. Reactorului 30, Bucharest-M\u{a}gurele 077125, Romania}
 \author{M. Karny} 
    \affiliation{Faculty of Physics, University of Warsaw, Warsaw, PL-02-093, Poland}  
 \author{L. H. Khiem} 
    \affiliation{Department of Nuclear Physics, Faculty of Physics, VNU University of Science, Hanoi, Vietnam }
 \author{G. G. Kiss } 
    \affiliation{RIKEN Nishina Center, Wako, Saitama 351-0198, Japan}  
    \affiliation{HUN-REN Institute for Nuclear Research, Debrecen Pf. 51, H-4001, Hungary}  
 \author{M. Kogimtzis }
    \affiliation{STFC Daresbury Laboratory, Daresbury, Warrington WA4 4AD, United Kingdom}
\author{A. Korgul} 
    \affiliation{Faculty of Physics, University of Warsaw, Warsaw, PL-02-093, Poland} 
 \author{S. Kubono } 
    \affiliation{RIKEN Nishina Center, Wako, Saitama 351-0198, Japan}  
 \author{M. Labiche } 
    \affiliation{STFC Daresbury Laboratory, Daresbury, Warrington WA4 4AD, United Kingdom}  
 \author{I. Lazarus } 
    \affiliation{STFC Daresbury Laboratory, Daresbury, Warrington WA4 4AD, United Kingdom}  
\author{J. Liang}
    \affiliation{Department of Physics and Astronomy, McMaster University, Hamilton, Ontario L8S 4-M1, Canada}
 \author{J. Lee } 
    \affiliation{Department of Physics, The University of Hong Kong, Pokfulam Road, Hong Kong, China}  
 \author{J. Liu } 
    \affiliation{Department of Physics, The University of Hong Kong, Pokfulam Road, Hong Kong, China}  
 \author{G. Lorusso } 
    \affiliation{RIKEN Nishina Center, Wako, Saitama 351-0198, Japan}  
    \affiliation{National Physical Laboratory, Teddington, UK TW11 0LW}  
 \author{K. Matsui } 
    \affiliation{RIKEN Nishina Center, Wako, Saitama 351-0198, Japan}   
    \affiliation{Department of Physics, University of Tokyo, Hongo, Bunkyo-ku, Tokyo 113-0033, Japan}  
 \author{K. Miernik } 
    \affiliation{Physics Division, Oak Ridge National Laboratory, Oak Ridge, TN 37831, USA}   
    \affiliation{Faculty of Physics, University of Warsaw, Warsaw, PL-02-093, Poland}  
 \author{F. Montes } 
    \affiliation{National Superconducting Cyclotron Laboratory, Michigan State University, East Lansing, MI 48824, USA}  
 \author{B. Moon } 
    \affiliation{Department of Physics, Korea University, Seoul 136-701, Republic of Korea}  
    \affiliation{Center for Exotic Nuclear Studies, Institute for Basic Science, Daejeon 34126, Republic of Korea}
 \author{A.I. Morales } 
    \affiliation{Instituto de Fisica Corpuscular (CSIC-Universitat de Valencia), E-46980 Paterna, Spain}  
 \author{N. Nepal } 
    \affiliation{Central Michigan University, Mount Pleasant, MI 48859, USA}  
 \author{S. Nishimura } 
    \affiliation{RIKEN Nishina Center, Wako, Saitama 351-0198, Japan}  
 \author{R. D. Page } 
    \affiliation{Department of Physics, University of Liverpool, Liverpool, L69 7ZE, United Kingdom}  
\author{M. Piersa-Si\l{}kowska} 
    \affiliation{Faculty of Physics, University of Warsaw, Warsaw, PL-02-093, Poland}
 \author{V. H. Phong } 
    \affiliation{RIKEN Nishina Center, Wako, Saitama 351-0198, Japan} 
    \affiliation{Department of Nuclear Physics, Faculty of Physics, VNU University of Science, Hanoi, Vietnam }
 \author{Zs. Podoly\'ak } 
    \affiliation{Department of Physics, University of Surrey, Guildford GU2 7XH, United Kingdom}  
 \author{V. F. E. Pucknell } 
    \affiliation{STFC Daresbury Laboratory, Daresbury, Warrington WA4 4AD, United Kingdom}  
 \author{P. H. Regan } 
    \affiliation{Department of Physics, University of Surrey, Guildford GU2 7XH, United Kingdom}  
    \affiliation{National Physical Laboratory, Teddington, UK TW11 0LW}  
 \author{B. Rubio } 
    \affiliation{Instituto de Fisica Corpuscular (CSIC-Universitat de Valencia), E-46980 Paterna, Spain}  
 \author{Y. Saito } 
    \affiliation{TRIUMF, 4004 Wesbrook Mall, Vancouver, BC, V6T 2A3, Canada}  
 \author{H. Sakurai } 
    \affiliation{RIKEN Nishina Center, Wako, Saitama 351-0198, Japan}   
    \affiliation{Department of Physics, University of Tokyo, Hongo, Bunkyo-ku, Tokyo 113-0033, Japan}  
 \author{Y. Shimizu } 
    \affiliation{RIKEN Nishina Center, Wako, Saitama 351-0198, Japan}  
 \author{J. Simpson } 
    \affiliation{STFC Daresbury Laboratory, Daresbury, Warrington WA4 4AD, United Kingdom}  
 \author{P.-A. S\"oderstr\"om } 
    \affiliation{RIKEN Nishina Center, Wako, Saitama 351-0198, Japan} 
    \affiliation{Extreme Light Infrastructure-Nuclear Physics (ELI-NP)/Horia Hulubei National Institute for Physics and Nuclear Engineering (IFIN-HH), Str. Reactorului 30, Bucharest-M\u{a}gurele 077125, Romania}
 \author{D. W. Stracener} 
    \affiliation{Physics Division, Oak Ridge National Laboratory, Oak Ridge, TN 37831, USA}  
 \author{T. Sumikama } 
    \affiliation{RIKEN Nishina Center, Wako, Saitama 351-0198, Japan} 
 \author{R. Surman}
    \affiliation{Department of Physics, University of Notre Dame, Notre Dame, Indiana 46656, USA}
 \author{H. Suzuki } 
    \affiliation{RIKEN Nishina Center, Wako, Saitama 351-0198, Japan}  
 \author{M. Takechi } 
    \affiliation{Department of Physics, Niigata University, Niigata 950-2102, Japan}             
 \author{H. Takeda } 
    \affiliation{RIKEN Nishina Center, Wako, Saitama 351-0198, Japan}  
 \author{A. Tarife\~no-Saldivia } 
    \affiliation{Universitat Politecnica de Catalunya (UPC), Barcelona, Spain}  
 \author{S. L. Thomas } 
    \affiliation{STFC Rutherford Appleton Laboratory, Harwell Campus, Didcot, Oxfordshire, OX11 0QX, UK}  
\author{M. Wolińska-Cichocka} 
    \affiliation{Heavy Ion Laboratory, University of Warsaw, PL-02-093 Warsaw, Poland} 
 \author{P. J. Woods } 
    \affiliation{School of Physics and Astronomy, The University of Edinburgh, Edinburgh, EH9 3FD, United Kingdom}  
 \author{X. X. Xu} 
    \affiliation{Department of Physics, The University of Hong Kong, Pokfulam Road, Hong Kong, China}  

\date{\today}

\begin{abstract}
Neutron emission probabilities and half-lives of 37 $\beta$\nobreakdash-delayed neutron emitters from \myisoSimp{75}{Ni} to \myisoSimp{92}{Br} were measured at the RIKEN Nishina Center in Japan, including 11 one-neutron and 13 two-neutron emission probabilities and 6 half-lives for the first time that supersede theoretical estimates. These nuclei lie in the path of the weak \rprocess{} occurring in neutrino-driven winds from the accretion disk formed after the merger of two neutron stars synthesizing elements in the $A\sim80$ abundance peak. The presence of such elements dominates the accompanying kilonova emission over the first few days and have been identified in the AT2017gfo event, associated to the gravitational wave detection GW170817. 
Abundance calculations based on over 17000 simulated trajectories describing the evolution of matter properties in the merger outflows show that the new data lead to an increase of 50\nobreakdash-70\% in the abundance of Y, Zr, Nb and Mo. This enhancement is large compared to the scatter of relative abundances observed in old very metal poor stars and thus is significant in the comparison with other possible astrophysical processes contributing to the light-element production.
These results underline the importance of including experimental decay data for very neutron-rich $\beta$\nobreakdash-delayed neutron emitters into $r$\nobreakdash-process models.
\end{abstract}

\maketitle

The origin of the chemical elements in the universe is a long standing open question~\cite{Arcones2023}. In particular, the astrophysical conditions needed for the synthesis of elements heavier than iron,  mainly driven by neutron capture processes \cite{b2fh,cameron57}, are under intense discussion. The so-called slow neutron capture process (\sprocess{}) happens when the neutron density is low and the time scale of neutron captures is similar to the subsequent $\beta$ decays. Therefore, this process runs close to stability and its contribution to the observed abundances is generally assumed to be well estimated \cite{kappeler2011}. Much more uncertain is the case of the rapid neutron capture process (\rprocess{}) that requires high neutron density and an explosive environment \cite{Horowitz19, Cowan21}. These conditions can be achieved in the matter ejected from compact binary mergers \cite{Lattimer.Schramm:1974, Freiburghaus_1999, Metzger10} and rare core-collapse supernovae with strong magnetic fields \cite{Winteler.etal:2012, Nishimura.etal:2017, Moesta.etal:2018, Reichert.etal:2021} (including the collapsar phase after the explosion \cite{Surman.etal:2006, Siegel.etal:2019, Miller.etal:2020}). 

The \rprocess{} has been observed already as an electromagnetic counterpart \cite{Cowperthwaite.etal:2017, Drout.etal:2017}, the AT2017gfo kilonova, to the gravitational wave detection of the neutron star merger event GW170817 \cite{Abbott.etal:2017ApJL, Abbott.etal:2017PRL}. Previous evidence of kilonovae were associated with gamma ray burst observations (see for example \cite{Tanvir13, Yang15}, and the review \cite{Metzger2019}) following early calculations of electromagnetic emission from compact binary mergers \cite{Li98}. The radioactive decay of the neutron-rich nuclei synthesized during the \rprocess{} produces the energy that is radiated away by the kilonova \cite{Metzger10, Metzger2019, Hotokezaka_2020, Zhu_2021}. Moreover, observations of the kilonova spectrum have confirmed the presence of freshly produced strontium ($Z$=38) in the matter ejected \cite{Watson2019}, as well as providing evidence for the production of yttrium ($Z$=39) and zirconium ($Z$=40) ~\cite{Gillanders2022,Domoto2022,Sneppen2023,Vieira2023}. All three elements are situated at the tail of the first abundance peak $A \sim 80$ associated with enhanced element production occurring while the $r$\nobreakdash-process path runs along the magic neutron number $N=50$~\cite{suess56}. The presence of first peak elements dominates the kilonova emission over the first few days~\cite{Kasen2017}, and a quantitative understanding of measured spectra, in this and future kilonovae, sheds light on the complex merger dynamics~\cite{Metzger2019}. Such understanding can be gained by comparison to detailed hydrodynamic simulations of the merger process and the inclusion of accurate nuclear physics input~\cite{Vieira2023}.

In our solar system Sr-Y-Zr are mainly the result of \sprocess{} nucleosynthesis~\cite{Arlandini1999,Bisterzo2014} although uncertainty in the exact proportion exists~\cite{Travaglio2004,Trippella2014}. To account for the remaining observed abundances apart from the \rprocess{} other mechanisms are considered, including an extended $\alpha$~process (or charged-particle reaction or weak \rprocess{}) \cite{Woosley92, Bliss18} and the intermediate neutron capture process ($i$~process), with neutron densities in between the $s$ and $r$~processes \cite{Cowan77,Cote2018}. 
Discerning among the abundance pattern of the different processes is best done looking into element abundances of old very metal poor (VMP) stars~\cite{Beers2005,Frebel:2018}, 
which are expected to collect the material from very few (ideally one) primary nucleosynthesis events. In this context it is worth mentioning that due to favorable spectroscopy conditions Sr, Y and Zr are readily observed in very old stars as can be corroborated by a glance to databases containing elemental abundances, like SAGA~\cite{Suda2018} and JINAbase~\cite{Abohalima2018}.

Comparison of observed abundances in selected groups of stars, enriched in $r$\nobreakdash-process material, shows that there is a small star-to-star scatter between heavy element ($Z > 55$) relative abundances, and also with respect to the solar $r$\nobreakdash-process fraction~\cite{sneden08,Ji2016}. This universality is interpreted as the signature of a $robust$ \rprocess{}. The situation for light elements is less clear and seems to depend on the selected star sample~\cite{Cowan21}. 
However, regularities in the abundance pattern of light elements can be found~\cite{Roederer2022} when normalized internally (among light elements). This supports the existence of additional scenarios with an enhanced production of light elements, and the attempts to disentangle empirically both components in individual stars~\cite{Montes.etal:2007, Hansen.etal:2014}.

Another approach is to compare observations to nucleosynthesis calculations based on simulations of the astrophysical sites.
These simulations are affected by large systematic uncertainties, coming from modelling or the required physics input. However, this does not diminish the importance of investigating the sensitivity of calculated abundances to nuclear data~\cite{Mumpower016}, which ought to be performed for each specific astrophysical model~\cite{Horowitz19}. Disentangling astrophysical and nuclear uncertainties is key to solving the heavy-element formation puzzle. Improving the accuracy of key nuclear data currently drives experimental and theoretical efforts world-wide~\cite{Horowitz19,Kajino2019,Cowan21}. In particular this is true for $\beta$\nobreakdash-decay data of neutron\nobreakdash-rich nuclei since the \rprocess{} runs far from the valley of $\beta$\nobreakdash-decay stability and measured data are scarce. Consequently one must resort to rather uncertain theoretical estimates (see for example recent state-of-the-art calculations in \cite{moller19,minato21,minato22}).

Beta decays are critical during different phases of the \rprocess{}. Initially, the neutron density and temperature are high enough and the \rprocess{} runs far from stability along very neutron-rich nuclei in an equilibrium between neutron capture and photodissociation, $(n,\gamma)-(\gamma, n)$ equilibrium. Beta decay allows escaping from equilibrium creating the next element, and thus its rate determines the speed at which heavier nuclei are produced.  Once most of the free neutrons have been captured, the remaining neutron-rich nuclei $\beta$ decay towards the valley of stability. The dominant decay branch of very neutron-rich nuclei is $\beta$\nobreakdash-delayed neutron emission. This channel opens whenever the neutron separation energy (\sn{}) in the daughter nucleus is lower than the available decay energy window (\qbeta{}). Population of states above \sn{} leads to the preferential emission of neutrons rather than $\gamma$-rays, characterized by the probability of one-, two-, ..., $x$-neutron emission ($P_{xn}$). Beta-delayed neutron emission influences the final abundances, altering the decay path and injecting fresh neutrons for late-time captures \cite{arcones11, Mumpower016}. 

In this work, we present the impact on nucleosynthesis of half-lives (\halflife{}) and one- and two-neutron emission probabilities ($P_{1n}$ and ($P_{2n}$)), of 37 newly measured neutron-rich nuclei close to and including doubly magic \myisoSimp{78}{Ni}. These nuclei span elements from Ni to Br in the mass range 75 to 93, many of them measured for the first time, or with better precision. More specifically we show their influence on elemental abundances obtained from a simulation of neutrino-driven wind ejecta after merging of two neutron stars~\cite{Perego2014,Dirk15} that peak around mass number $A \sim 80$ and can contribute with up to 1\% of solar mass ($M_{\odot}$) to the nucleosynthesis yield.

The measurement was among the goals of the experimental campaign carried out by the BRIKEN Collaboration \cite{Tain18} at the RIKEN Nishina Center. Results were already published for nuclei contributing to the second \rprocess{} abundance peak \cite{ohall19, Phong2022} and the rare-earth peak \cite{Kiss2022}. Here we report on global results for nuclei around \myisoSimp{78}{Ni} adding on those for gallium isotopes, already presented in \cite{Yokoyama2019,Yokoyama2023} where the analysis method described in \cite{charlie19} is used. We summarize below the measurements and results from the present work. Details on the instrumentation and methods are given in \cite{TolosaDelgado19}. Further details of the analysis and results can be found in \cite{atd_thesis}.

\begin{figure}[htb]
\centering
\includegraphics[width=\linewidth]{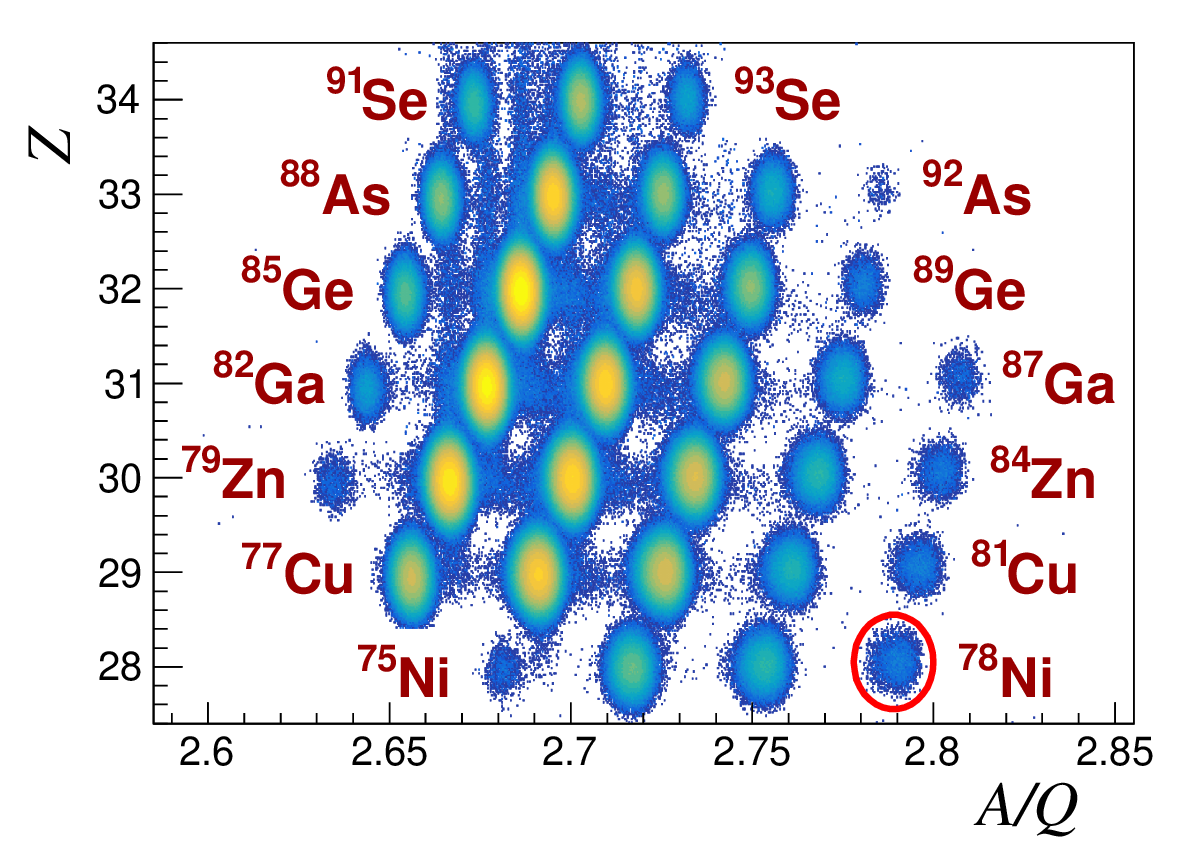} 
\caption{Nuclide identification of the implanted ions accumulated during the experiment RIBF127. The horizontal axis corresponds to the mass-over-charge ratio ($A/Q$) and the vertical axis to the atomic number ($Z$). The labels indicate the range of isotopes investigated.
Data for \myisoSimp{75}{Ni}, \myisoSimp{76}{Cu}, \myisoSimp{79}{Zn}, and \myisoSimp{82}{Ga} come from the commissioning run \cite{TolosaDelgado19}. Results for \myisoSimp{87}{As} and \myisoSimp{92}{Br} are derived from parent data.
The red circle highlights the doubly-magic nucleus \myisoSimp{78}{Ni} with $\sim7600$ implanted ions.}
\label{fig:figPID}
\end{figure}

The experimental setup consists of the Advanced Implantation Detector Array (AIDA) \cite{Hall2023}, surrounded by the highly efficient BRIKEN neutron counter \cite{Tarife_o_Saldivia_2017,TolosaDelgado19}. The radioactive beam is produced by in-flight fission of a high-intensity ($\sim30$ pnA) $^{238}\textrm{U}$ beam at 345~MeV/nucleon impinging on a 4~mm thick beryllium target. The resulting cocktail of radioactive nuclei traverses the BigRIPS and ZeroDegree Spectrometer \cite{kubo12}, which separate and identify nuclides using the $TOF-B\rho-\Delta E$ method \cite{fukuda13}. Finally nuclei of interest are stopped in AIDA. Figure~\ref{fig:figPID} displays the nuclide identification plot of the implanted ions accumulated during the RIBF127 experiment. The implantation of each radioactive nucleus and its subsequent decays are detected by one of the six highly segmented Double-sided Silicon Strip Detectors (DSSD) of AIDA. Decays accompanied by the emission of one or more neutrons are detected by the array of 140 $^3\textrm{He}$ proportional tubes of the BRIKEN neutron counter.

Half-lives and neutron emission probabilities for each isotope are obtained from the analysis of suitable decay curves. These are generated from the time difference between decays and identified implants in AIDA. In order to reduce the random implant-decay time correlations, we add the condition  that implant and decay position pixels in the detector must overlap. The one-neutron emission and two-neutron emission decay curves are constructed adding the proper coincidence conditions with the neutron events. The emission of more than two neutrons was not observed for nuclei studied in this work within our sensitivity, in accordance with theoretical estimates \cite{moller19,Marketin2016}.

The half-life, the neutron emission probabilities of the parent nuclei and the number of parent decays are obtained by simultaneously fitting the $\beta$\nobreakdash-decay, one-neutron $\beta$\nobreakdash-decay and two-neutron $\beta$\nobreakdash-decay curves using the Binned Maximum Likelihood method \cite{ROOT_Fit}. Half-lives and neutron emission probabilities of descendant nuclei are kept fixed during the fit and taken from the most recent evaluation fostered by the IAEA \cite{Birch15,Liang20}. Details about the fitting functions and a novel method to evaluate the correlated $\beta$\nobreakdash-neutron background are given in \cite{TolosaDelgado19}. 

The energy dependence of neutron detection efficiency introduces a systematic uncertainty in the result, since the neutron emission spectrum is generally unknown. We developed a method, summarized in Appendix A of End Matter section, to assign an efficiency and uncertainty for each nucleus.
Additional systematic uncertainties correspond to uncertainties in \halflife{} and \pxn{} of descendants, and the amount of uncorrelated and correlated backgrounds. To evaluate their contribution we use a Monte~Carlo method, sampling repeatedly all parameters before re-fitting. The final uncertainty of the results is calculated by adding in quadrature the systematic uncertainty to the statistical uncertainty provided by the fitter.

\begin{figure}[!ht]
\includegraphics[width=\linewidth]{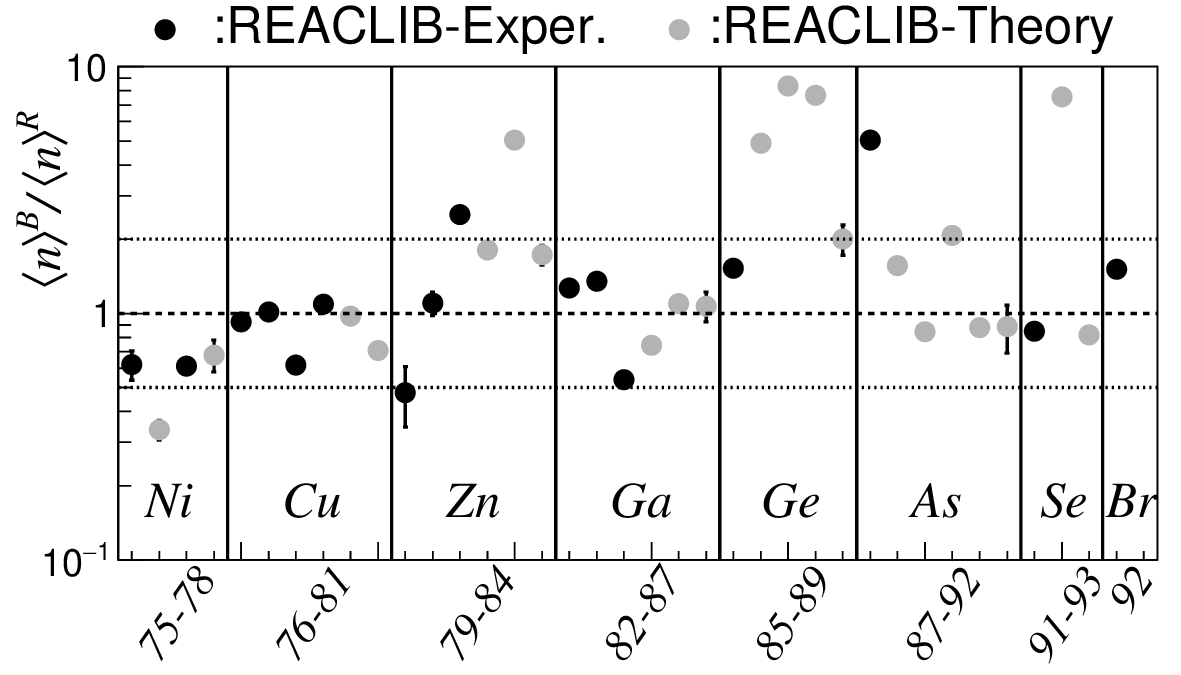}
\includegraphics[width=\linewidth]{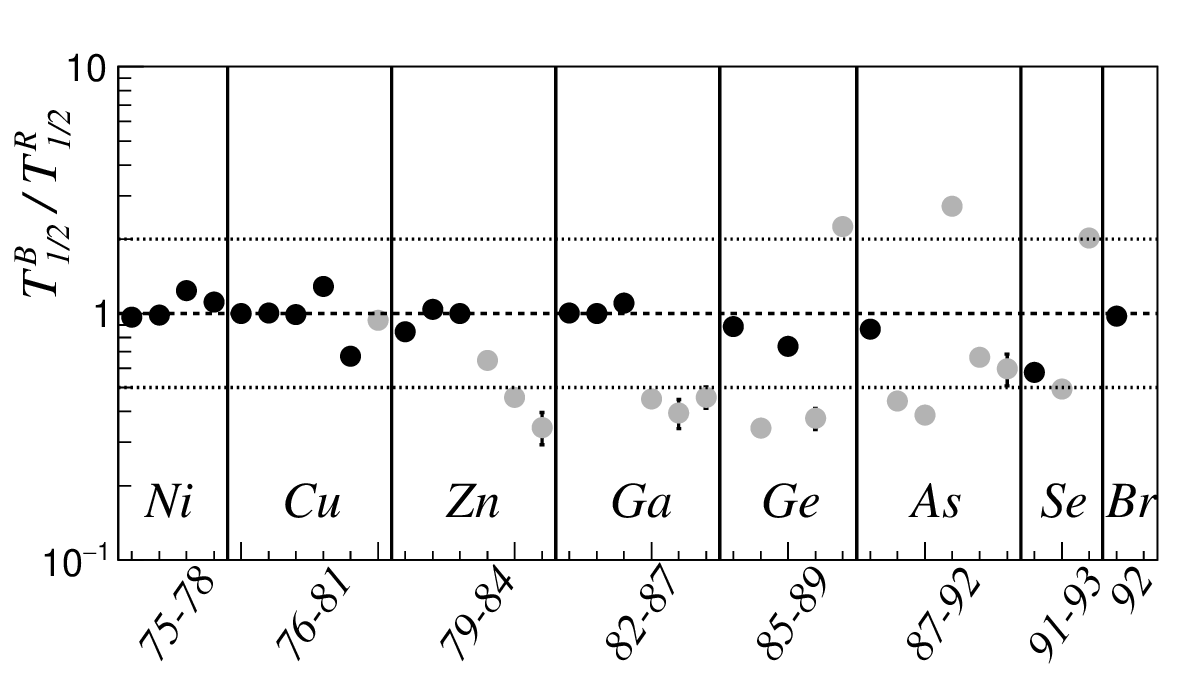}
\caption{Upper panel: ratio of the average number of neutrons emitted per decay  obtained in this work, $\langle n \rangle^{B}$, and the values derived from the JINA REACLIB database, $\langle n \rangle^{R}$. Lower panel: ratio of half-lives values obtained in this work, $T^{B}_{1/2}$, and the values included in REACLIB, $T^{R}_{1/2}$. Black and grey symbols correspond to experimental and theoretical REACLIB values respectively. The error bars represent the uncertainty in our data. As a reference, the two dotted horizontal lines represent ratios of 2 and $1/2$.}
\label{fig:compreaclib}
\end{figure}

As a result we determined for the first time 11 \ponen{} values (\myisoSimp{87-89}{Ge}, \myisoSimp{88-92}{As}, \myisoSimp{92,93}{Se}), including the doubly-magic nucleus \myisoSimp{78}{Ni}, and 13 \ptwon{} values (\myisoSimp{80,81}{Cu}, \myisoSimp{83,84}{Zn}, \myisoSimp{87-89}{Ge}, \myisoSimp{88-92}{As}, \myisoSimp{93}{Se}). We re-analyzed the data for gallium isotopes and the results for \ponen{} and \ptwon{} are consistent within uncertainties with those presented in \cite{Yokoyama2023}. We improved the uncertainty for the remaining measured cases, 26 \ponen{} and 2 \ptwon{} values, when compared to the IAEA evaluation \cite{Birch15,Liang20}.  These include the \ponen{} values of \myisoSimp{87}{As} and \myisoSimp{92}{Br} that were obtained from the fit of parent \myisoSimp{87}{Ge} and \myisoSimp{92}{Se} decay curves. We obtained 7 new \halflife{} values (\myisoSimp{89}{Ge}, \myisoSimp{89-92}{As}, \myisoSimp{92,93}{Se}), improved the uncertainty for 6 nuclei (\myisoSimp{80,81}{Cu}, \myisoSimp{79}{Zn}, \myisoSimp{87}{Ga} \myisoSimp{88}{As} and \myisoSimp{91}{Se}) and confirmed the values of the evaluation \cite{Birch15,Liang20} for the remaining cases. Data for \myisoSimp{75}{Ni}, \myisoSimp{76}{Cu}, \myisoSimp{79}{Zn}, and \myisoSimp{82}{Ga} come from the BRIKEN commissioning run \cite{TolosaDelgado19}, prior to RIBF127 experiment, which had better statistics. A Table with numerical values is included in Appendix B of End Matter section. A comparison of our data with previous data and theoretical predictions is shown in Fig.~\ref{fig:compreaclib}, as explained below.

In order to perform nucleosynthesis calculations a two-step approach is followed. In the first step, the time evolution of relevant properties of the matter ejected in the selected $r$\nobreakdash-process scenario is obtained from three-dimensional hydrodynamic simulations. Tracer particles (also called trajectories or mass elements) are included in the simulations to track the evolution of density, temperature, electron fraction, and neutrino fluxes and energies. In this work we consider a simulation  of a neutron star merger where a massive neutron star is formed and lives for 190~ms~\cite{Dirk15,Perego2014} before collapsing into a black-hole. The neutrino wind heats and ejects matter ($\sim 9 \times 10^{-3} M_{\odot}$) within a polar angle of about $60^{\circ}$ because of the shading effect of the equatorial accretion disk. The simulation shows that the evolution of relevant medium properties depends on ejection angle, thus it is essential to perform the calculations for the full ensemble of trajectories in order to evaluate the impact of the new decay data. From the initial $10^{5}$ tracers, over $17\times10^{3}$ trajectories that become gravitationally unbound are stored for further processing.
The electron fraction ($Y_{e}$) of the matter ejected varies in the range $0.2-0.4$ and peaks at $\sim 0.32$. In these conditions a weak \rprocess{} takes place, i.e. a neutron\nobreakdash-capture driven process with seed nuclei in the mass range $A \sim 50-85$ and a low neutron-to-seed ratio, which is a candidate for the light element ($A<140$) production mechanism observed in old VMP stars.

In a second step (post-processing), the tracer information for the full set of unbound trajectories is combined with nuclear physics data to serve as input for a nuclear reaction network. The WinNet network \cite{Winteler_2012,Reichert2023} was used to calculate the evolution of the abundances of more than 5000 nuclei over time for each trajectory. This calculation requires a huge amount of reaction and decay data, which are mainly compiled in the JINA REACLIB database \cite{reaclib}. This database includes, among other parameters, $\beta$\nobreakdash-decay half-lives and neutron-emission branching ratios \pxn{}, taking experimental values whenever possible from \cite{nwc11}. Otherwise, theoretical \halflife{} values from \cite{mo03pn} are used, while theoretical neutron emission probabilities are taken from \cite{mum16}. The REACLIB database is complemented with additional reaction rates as described in \cite{Reichert2023}.

In order to  explore the  impact of our newly measured \halflife{} and \pxn{}, two nuclear input databases are combined with the full ensemble of trajectories. One of the nuclear databases contains the original REACLIB values, and the other one is updated with our new experimental values. Since the \rprocess{} runs away from stability, neutron emission probabilities and half-lives of many of the involved nuclides come from theoretical estimates. Our data replaces calculated values of neutron emission for 21 nuclei and of half-life for 17 nuclei. We observe sizeable differences with respect to some experimental values in the database.

The upper panel in Fig.~\ref{fig:compreaclib} shows a comparison, as ratios, of the average number of neutrons emitted per decay $\langle n \rangle$ (calculated as \ponen{} + 2\ptwon{}), derived from our experimental values and the values in the REACLIB database. Our values differ by over 30\% with respect to REACLIB for 10 of the nuclei with experimental value and for 13 nuclei with theoretical estimate. Deviations in excess of a factor 5 are observed for \myisoSimp{87}{As} (experimental) and \myisoSimp{83}{Zn}, \myisoSimp{86-88}{Ge} and \myisoSimp{92}{Se} (theoretical). There is no obvious $Z$ or $A$ pattern in the discrepancies with theoretical values, but on average they are too low compared to experiment. These estimates are obtained~\cite{mum16} from Gamow-Teller (GT) $\beta$\nobreakdash-decay strength distributions based on wave-functions calculated within the quasi-particle random phase approximation (QRPA)~\cite{Moeller1997}. As an improvement over prior works, $P_{xn}$ values in \cite{mum16} take into account the competition between the different $xn$ and $\gamma$ emission channels using the statistical Hauser-Feshbach theory. 

The bottom panel in Fig.~\ref{fig:compreaclib} displays the comparison of our experimental half-lives with those included in the REACLIB database. Experimental half-lives included in REACLIB agree well with the results from the IAEA evaluation \cite{Birch15,Liang20}. Compared to experimental data in the REACLIB our half-lives differ less than 30\% for all nuclei except for \myisoSimp{80}{Cu} and \myisoSimp{91}{Se}, which are respectively 1.5 and 1.7 times larger in the database. When compared with theoretical estimates in the REACLIB, our values, except for \myisoSimp{81}{Cu}, differ by more than 30\%, most of them showing ratios grouping around a factor of 2 or $1/2$. As in the case of neutron yields there is not an obvious pattern in the deviations, but on average the estimated values are too high compared to our experimental values. The calculated $T_{1/2}$~\cite{mo03pn} are also based on QRPA wave functions, but in this case the $\beta$\nobreakdash-decay strength includes both Gamow-Teller and first-forbidden transitions, aiming to improve the predictions in particular near closed neutron shells. 

\begin{figure}[!ht]
\includegraphics[width=\linewidth]{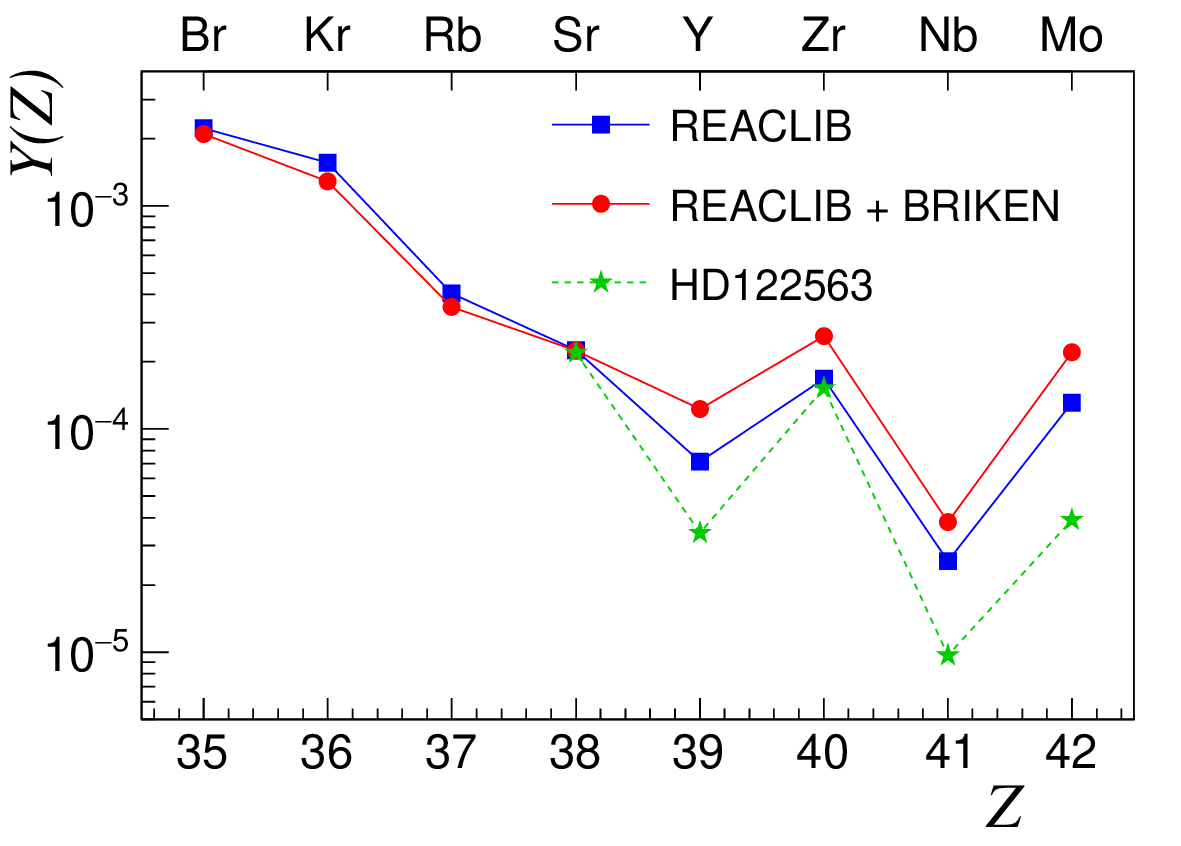}
\caption{Calculated abundances using the default REACLIB database  (blue squares) or the same database updated with the experimental values in this work REACLIB+BRIKEN (red circles). The elemental abundance is calculated as $Y(Z) = \sum_{A} Y(Z,A)$, $Y(Z,A)$ being defined as the ratio of the number of \myisoSimp{A}{Z} nuclei to the number of nucleons for all nuclei in a given volume~\cite{Reichert2023}. Abundances from light \rprocess{} enhanced star HD122563~\cite{Honda2007} normalized to Sr are also shown for reference (green stars).}
\label{fig:calc_abund}
\end{figure}

The impact of our data on the nucleosynthesis calculation is shown in Fig.~\ref{fig:calc_abund} displaying the abundances as a function of atomic number for the original REACLIB database and for the updated database (REACLIB+BRIKEN). In both cases the abundance peaks at $Z \sim 34$ (outside the range shown). We observe that the new data have no or little impact on Sr and elements below but have a sizeable effect on elements up to Mo, increasing the abundance of Y and Mo by $\sim 70$\% and that of Zr and Nb by $\sim 50$\%. 
As shown in Appendix C of End Matter section, no single $\beta$-decay exerts a dominant effect, and both \halflife{} and \pxn{} are important.

The differences for Y-Mo are significantly larger than the spread found in \cite{Roederer2022} for light element abundance ratios in VMP stars, and are therefore relevant in the discussion of the contribution of different processes to the observed abundances. The authors of \cite{Roederer2022} obtain a median absolute dispersion (MAD) smaller than 17\%  on the relative to Zr abundances of Sr, Y, Nb and Mo for 8 selected stars and values of 28\% (Sr), 23\% (Y) and 17\% (Mo) for a much larger sample of stars from JINAbase~\cite{Abohalima2018}, consisting of 294, 294 and 13 objects respectively. We include in Fig.~\ref{fig:calc_abund} for reference the abundance distribution observed~\cite{Honda2007} in the VMP star HD~122563, representative of stars enriched in light $r$\nobreakdash-process elements. 
We should emphasize that the impact of our data in the abundances obtained in the current simulation is robust, since it is the average of the full ensemble of trajectories and it reflects the fact that most of the corresponding $r$\nobreakdash-process paths run over the nuclei investigated in this work. We observed for another astrophysical site investigated~\cite{atd_thesis}, the ejecta from a neutron star merger accretion disk~\cite{Rodri17}, that selecting a few representative trajectories can lead to a misleading interpretation about the importance of nuclear data  since they can be washed out in the full calculation. Such a warning may also apply to parametric studies of nuclear data sensitivity in nucleosynthesis calculations, where the complexity of the astrophysical event is captured in a few global parameters.

In summary, we have evaluated the astrophysical impact of new experimental \halflife{} and \pxn{} values of 37 $\beta$\nobreakdash-delayed neutron emitters around the doubly-magic nucleus \myisoSimp{78}{Ni} in the mass range $A=75-93$. These nuclei lie on the path of the \rprocess{} occurring in neutrino-driven winds ejected from disks formed after a neutron-star merger and have an effect on the abundances of elements in the tail of the first \rprocess{} peak. In particular, compared to the base calculation, the production of Y is increased by $\sim 70\%$ and of Zr by $\sim 50\%$ while that of Sr is unchanged. These elements have been observed in the 2017 kilonova event and are readily identified as light \rprocess{} elements in VMP stars with a significantly smaller relative abundance scatter. Therefore the changes are relevant when trying to disentangle the contribution of different scenarios to the observed abundances in light element enhanced stars.

Simulations of the \rprocess{} require a vast amount of nuclear physics data as input, provided by experiments and theoretical models. Our results show that new experimental data are required to improve these models, and increase the reliability of simulations, so they can be used to narrow down the astrophysical site of the \rprocess{} and ultimately to understand the weight of each astrophysical scenario in the galactic chemical evolution throughout the history of the Universe. 

\vspace{0.5cm}

This experiment was performed at the RI Beam Factory operated by RIKEN Nishina Center and CNS, University of Tokyo. Work supported by Spanish grants MINECO FPA2014-52823-C2-1-P, MICINN/AEI FPA2017-83946-C2-1-P, and MICINN/AEI PID2019-104714GB-C21, co-funded by the European Regional Development Fund. Partially supported by Generalitat Valenciana PROMETEO/2019/007. AA was supported by the Deutsche Forschungsgemeinschaft (DFG, German Research Foundation) – Project-ID 279384907 - SFB 1245 and by the State of Hesse within the Research Cluster ELEMENTS (Project ID 500/10.006). A. T. D. acknowledges support from pre-doctoral fellowship BES-2015-075066, and Academy of Finland. A.I.M. work was supported by Generalitat Valenciana (CISEJI/2022/25), Ministerio de Ciencia, Innovación y Universidades (CAS22/00114 and CNS2023-144871). This work has been supported by the Office of Nuclear Physics, U. S. Department of Energy under the contract DE-AC05-00OR22725 (ORNL). I. D. and R. C. F. were supported by the Canadian NSERC grant SAPIN 2014-00028 and 2019-00030. M. R. acknowledges support from the Juan de la Cierva programme (FJC2021-046688-I) funded by MCIN/AEI/10.13039/501100011033 and by the European Union “NextGenerationEU". Supported from grant PID2021-127495NB-I00 and “ESF Investing in your future”. Additionally, he acknowledges support from the Astrophysics and High Energy Physics programme and the Prometeo excellence programme of the Generalitat Valenciana (ASFAE/2022/026 and CIPROM/2022/13). G. K. K. work was supported by NKFIH (K147010). M. P. S. acknowledges the funding support from the Polish National Science Center under Grant No. 2020/36/T/ST2/00547. A. K. acknowledges the funding support from the Polish National Science Center under Grant No. 2020/39/B/ST2/02346. A. A.  acknowledges partial support of the JSPS Invitational Fellowships for Research in Japan (ID: L1955). S. B. acknowledges the funding support from the Korean National Research Foundation grants NRF-2015H1A2A103027 and NRF-21A20131111123. B. M.  acknowledges the support from Institute for Basic Science in Republic of Korea (Grant No. IBS-R031-Y1). S. N acknowledges the JSPS KAKENHI (Grants No. 17H06090, 20H05648, 22H04946) and the RIKEN program  for Evolution of Matter in the Universe (r-EMU and RiNA-Net). UK authors were supported by the UK Science and Technology Facilities Council through grant numbers ST/P004598/1 and ST/V001027/1. D. A. work was supported by Institute for Basic Science (Grant No. IBS-R031-D1). P. A. S. acknowledge the contract PN 23.21.01.06 sponsored by the Romanian Ministry of Research, Innovation and Digitalization. 

\bibliography{main}

\clearpage

\section{Appendix A: Systematic uncertainty on nucleus-dependent neutron detection efficiency}

The dependence of neutron detection efficiency on energy for moderated neutron counters introduces a systematic uncertainty in the extracted \pxn{} since the emitted neutron energy spectrum is generally unknown.
For the BRIKEN counter the detection efficiency is determined \cite{TolosaDelgado19}, combining experimental information and Monte Carlo simulations, to be rather constant for neutrons with energy below 1~MeV  at a value of 66.8(20)\%, but it decreases for higher neutron energies reaching, for example, 58.5\% (simulated) detection efficiency for 3~MeV neutrons \cite{Tarife_o_Saldivia_2017}. Nuclei studied in this work have $\beta$\nobreakdash-decay energy windows for neutron emission, $Q_{\beta n} = Q_{\beta} - S_n$, reaching up to 14~MeV, thus a method was introduced \cite{atd_thesis} to estimate the efficiency and its uncertainty for each $\beta$\nobreakdash-delayed neutron emitter.

The starting point is a simplified model of $\beta$\nobreakdash-delayed neutron emission \cite{Pappas72}, based on a constant $\beta$\nobreakdash-decay strength $S_{\beta} (E_{x})$ above the neutron separation energy \sn{} and restricting neutron population to the ground state of the end nucleus. We further assume no dependency on spin-parity of the different states. All this is equivalent to saying that neutron spectra are just proportional to the statistical rate function $f (Z, Q_{\beta} - E_x)$ \cite{konopinski1965}, which carries the final-state Coulomb and phase space effect on $\beta$\nobreakdash-decay intensity. These seemingly drastic assumptions have the virtue of allowing easy computation (see Eq.~\ref{eq:avEn}) of the average neutron energy $\langle E_n \rangle$ as a function of neutron emission energy window $Q_{\beta n}$, sufficiently well known for all studied nuclei \cite{Wang_2021}, and of setting the proper scale to compare with experimental values. 

\begin{equation}
    \langle E_n \rangle = \frac{\int_{S_n}^{Q_{\beta}} f (Z, Q_{\beta} - E) E dE}{\int_{S_n}^{Q_{\beta}} f (Z, Q_{\beta} - E) dE}
    \label{eq:avEn}
\end{equation}

\begin{figure}[htb]
\centering
\includegraphics[width=\linewidth]{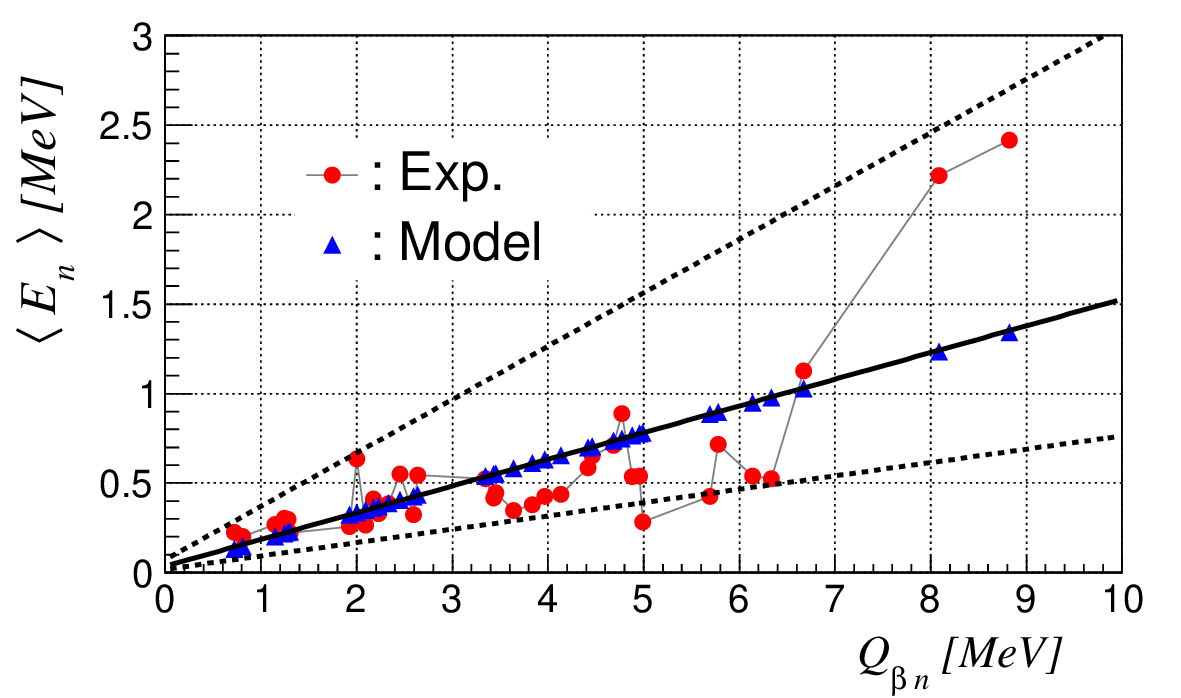} 
\caption{Average $\beta$-delayed neutron energy $\langle E_n \rangle$ as a function of emission window $Q_{\beta n}$. Red cycles represent experimental values obtained from measured spectra for medium-heavy nuclei \cite{Brady1989,Madurga2016}. The continuous black line is a linear fit to model values (blue triangles). Dashed black lines show the result of scaling  the fit line by a factor 2 and 0.5 respectively.}
\label{fig:figAvEnQbn}
\end{figure}

This is shown in Fig.~\ref{fig:figAvEnQbn} for the 36 medium-heavy delayed neutron emitters that have available experimental neutron spectra \cite{Brady1989,Madurga2016}. Note that average energies are always lower than 30\% of the emission energy window, showing the compression effect of the strong energy dependence of the $f$-function.

As seen in Fig.~\ref{fig:figAvEnQbn}, model average energies (blue triangles) follow very closely a linear dependency on $Q_{\beta n}$ (continuous black line) and the experimental values (red circles) scatter around, reflecting nuclear structure details in the strength function $S_{\beta} (E_{x})$. Maximum deviations are contained, except for one point, within a factor of 2 (dashed black lines) of the model value. Rather conservatively, we take this value as the uncertainty on the model $\langle E_n \rangle$.

The second assumption in our method is that the efficiency at the average neutron energy is a very good proxy for the average efficiency over the neutron spectra, $\epsilon (\langle E_n \rangle) \approx \langle \epsilon (E_n) \rangle$. Indeed we verified that for the 36 nuclei in Fig.~\ref{fig:figAvEnQbn} the difference is below 0.5\%. This allows an easy computation of neutron detection efficiency and uncertainty as a function of $Q_{\beta n}$ using the experimentally-benchmarked simulated efficiency. As a reference, for  $Q_{\beta n} = 11.5$~MeV (\myisoSimp{92}{As}) we used an efficiency of $63.5^{+3.8}_{-6.0}$~\%. We found that the uncertainty on neutron detection efficiency is in general a major contribution to the overall systematic uncertainty (see main text).

\section{Appendix B: Half-lives and neutron emission probabilities obtained in this work}

\setlength{\LTcapwidth}{0.5\textwidth}
\renewcommand{\arraystretch}{1.2}
\renewcommand{\doublerulesep}{0pt}
\begin{longtable}[e]{@{\extracolsep{\fill}}c r r r}
\hline\hline\hline
Nucleus & $T_{1/2}$ & $P_{1n}$ & $P_{2n}$ \\
\hline
\endfirsthead
\hline\hline\hline
Nucleus & $T_{1/2}$ & $P_{1n}$ & $P_{2n}$ \\
\hline
\endhead
\hline\hline
\endfoot
\hline\hline\\[1ex]
\caption{Half-lives $T_{1/2}$, one-neutron emission probabilities $P_{1n}$ and two-neutron emission probabilities $P_{2n}$ determined in this work. The quoted uncertainties include both statistical and systematic uncertainties (see main text). Half-life values marked with the symbol $^{\dagger}$ correspond to cases where the initial fit provide $T_{1/2}$ compatible with the value in the recent IAEA evaluation (see Ref.~\cite{Birch15,Liang20}) within the quoted uncertainty in the database, and did not improve that uncertainty. Subsequently the half-life was fixed to the database value, given in the table, in order to improve the accuracy of the $P_{xn}$ fit. \label{results}}
\endlastfoot
\myisoSimp{75}{Ni} & $331.8 (32)^{\dagger}$ & $6.20 (83)$ & $-$	\\
\myisoSimp{76}{Ni} & $234.7 (27)^{\dagger}$ & $8.78 (82)$ & $-$	\\
\myisoSimp{77}{Ni} & $158.4 (44)^{\dagger}$ & $18.4 (15)$ & $-$	\\
\myisoSimp{78}{Ni} & $122.2 (51)^{\dagger}$ & $25.8 (38)$ & $-$	\\
\myisoSimp{76}{Cu} & $637.5 (80)^{\dagger}$ & $6.65 (55)$ & $-$	\\
\myisoSimp{77}{Cu} & $469.8 (20)^{\dagger}$ & $30.8 (13)$ & $-$	\\
\myisoSimp{78}{Cu} & $331.7 (20)^{\dagger}$ & $40.2 (15)$ & $-$	\\
\myisoSimp{79}{Cu} & $241.3 (21)^{\dagger}$ & $59.9 (24)$ & $-$	\\
\myisoSimp{80}{Cu} & $114.0 (24)$ & $56.5 (28)$ &	$< 0.55$	\\
\myisoSimp{81}{Cu} & $75.8 (38)$ & $71.1 (57)$ & $2.2 (13)$	\\
\myisoSimp{79}{Zn} & $839.1 (75)$ & $0.62 (17)$ & $-$	\\
\myisoSimp{80}{Zn} & $562.0 (30)^{\dagger}$ & $1.10 (12)$ & $-$	\\
\myisoSimp{81}{Zn} & $303.7 (31)^{\dagger}$ & $18.88 (73)$ & $-$	\\
\myisoSimp{82}{Zn} & $177.9 (25)^{\dagger}$ & $61.4 (25)$ &	$-$	\\
\myisoSimp{83}{Zn} & $99.7 (30)^{\dagger}$ & $57.4 (29)$ & $2.17 (44)$	\\
\myisoSimp{84}{Zn} & $53.6 (81)^{\dagger}$ & $64.0 (59)$ &	$< 2$	\\
\myisoSimp{82}{Ga} & $600.9 (20)^{\dagger}$ & $25.2 (11)$ & $-$	\\
\myisoSimp{83}{Ga} & $308.4 (10)^{\dagger}$ & $84.7 (44)$ & $-$	\\
\myisoSimp{84}{Ga} & $92.70 (70)^{\dagger}$ & $37.3 (22)$ & $1.61 (13)$ \\
\myisoSimp{85}{Ga} & $91.9 (12)^{\dagger}$ & $71.2 (25)$ & $1.90 (16)$ \\
\myisoSimp{86}{Ga} & $50.3 (68)^{\dagger}$ & $57.6 (40)$ & $16.2 (19)$ \\
\myisoSimp{87}{Ga} & $31.3 (31)$ & $81 (11)$ & $15.3 (50)$ \\
\myisoSimp{85}{Ge} & $496.7 (60)^{\dagger}$ & $21.4 (14)$ & $-$	\\
\myisoSimp{86}{Ge} & $221.6 (11)^{\dagger}$ & $24.5 (12)$ & $-$	\\
\myisoSimp{87}{Ge} & $103.2 (35)^{\dagger}$ & $33.5 (15)$ & $0.26 (6)$	\\
\myisoSimp{88}{Ge} & $60.8 (61)^{\dagger}$ & $37.8 (27)$ & $0.22 (16)$	\\
\myisoSimp{89}{Ge} & $43.1 (24)$ & $43.9 (62)$ & $2.1	(13)$	\\
\myisoSimp{87}{As} & $484 (35)^{\dagger}$ &	$77.9 (39)$ & $-$ \\
\myisoSimp{88}{As} & $189.1 (25)$ & $53.1 (21)$ &	$0.48 (17)$	\\
\myisoSimp{89}{As} & $81.05 (71)$ & $83.8 (27)$ & $0.250 (50)$ \\
\myisoSimp{90}{As} & $54.4 (12)$ & $76.0 (37)$ & $5.46 (55)$ \\
\myisoSimp{91}{As} & $38.3 (16)$ & $84.1 (49)$ & $2.13 (61)$ \\
\myisoSimp{92}{As} & $18.3 (27)$ & $63 (16)$ & $23.0 (89)$ \\
\myisoSimp{91}{Se} & $155.6 (37)$ & $17.8	(14)$ &	$-$	\\
\myisoSimp{92}{Se} & $67.54 (72)$ & $15.14 (76)$ & $-$ \\
\myisoSimp{93}{Se} & $47.4 (20)$ & $24.7 (37)$ &	$0.79 (57)$	\\
\myisoSimp{92}{Br} & $334 (14)^{\dagger}$ &	$49.9 (18)$ & $-$ \\
 \end{longtable}

\section{Appendix C: Sensitivity of abundance calculations to individual and aggregated new nuclear data}

A question that naturally arises when assessing the impact of an ensemble of nuclear data on calculated elemental abundances is the relative importance of individual items. However, a calculation involving the full set of $17\times10^{3}$ trajectories used in this work is computationally very expensive, and the direct investigation of the impact of individual \halflife{} and \pxn{} and their correlations is not feasible.

To make such analysis more manageable, we reduced the number of tracer particles by binning them in electron fraction $Y_e$ space and selecting a reduced number of representative trajectories in each bin \cite{Reichert2023b}. Using a bin size of $\Delta Y_e = 0.05$ and selecting 5 tracers per bin, or less when not available, resulted in the selection of about 200 tracers. We verified that such a subset reproduces to a good approximation the abundances calculated with the original REACLIB database and with the updated REACLIB+BRIKEN database using the full set of tracers (Fig.~3 in the main text), while using $\sim80$ times less computation time.

\begin{figure}[t]
\centering
\includegraphics[width=\linewidth]{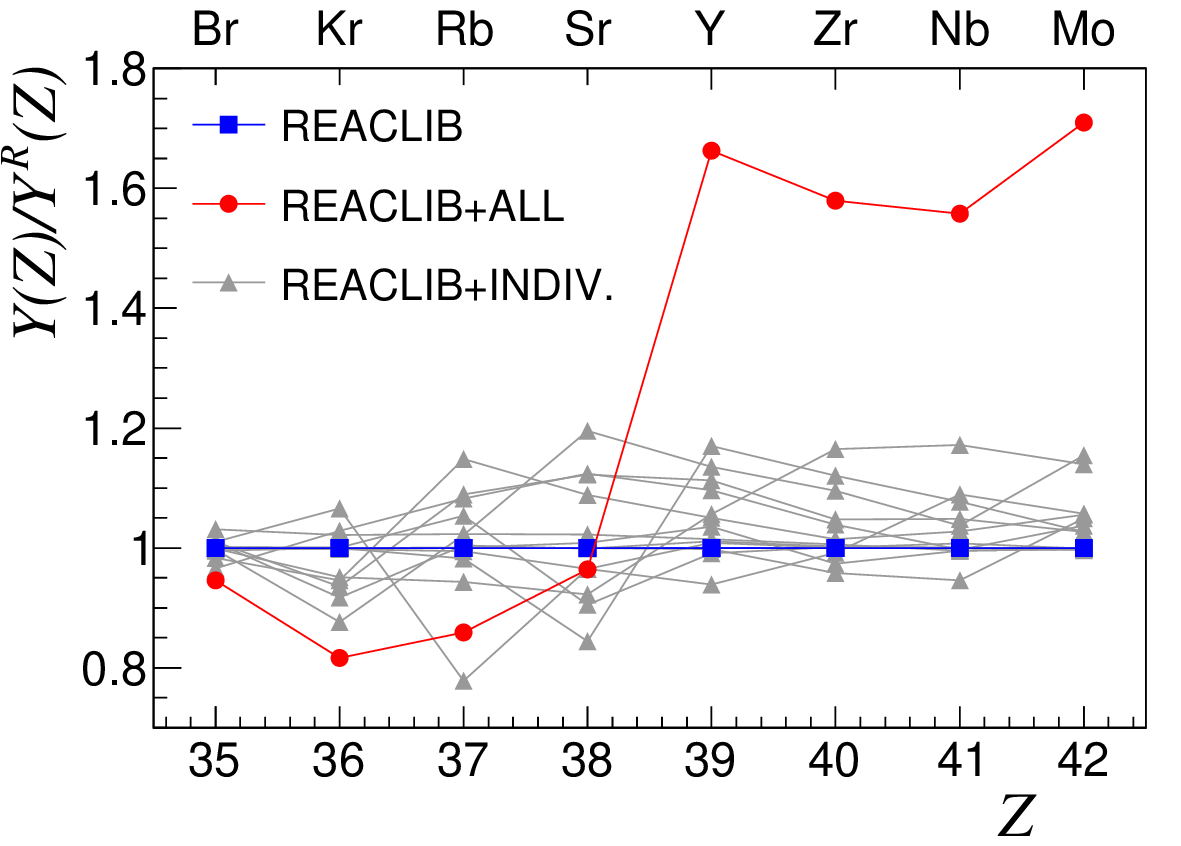} 
\includegraphics[width=\linewidth]{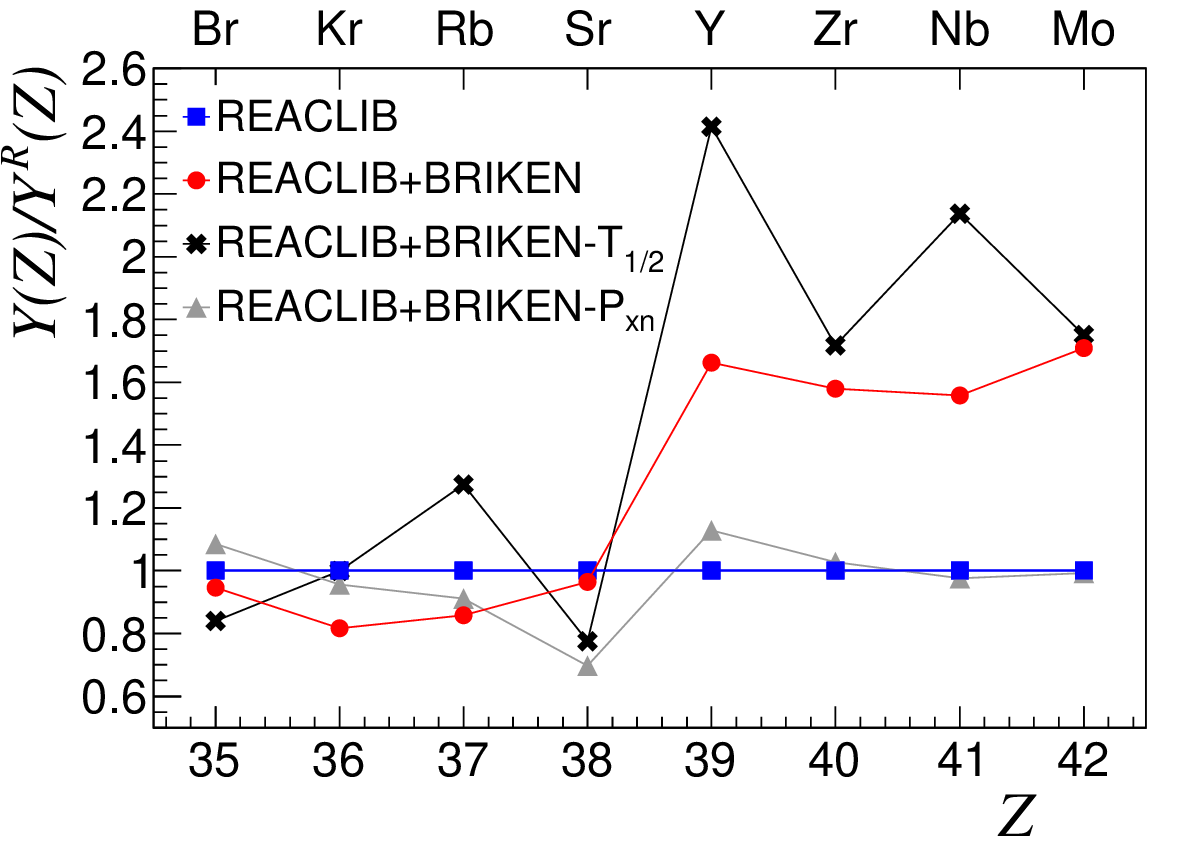} 
\caption{Abundances calculated with the reduced set of trajectories $Y(Z)$ normalized to the calculation with the unchanged REACLIB $Y^{R}(Z)$ (blue squares). The abundances calculated after replacing simultaneously  \halflife{} and \pxn{} for the 37 nuclei studied in this work are represented by red circles. \textit{Upper panel:} Gray triangles represent abundances calculated replacing both \halflife{} and \pxn{} for each of the 37 nuclei in turn. Only the 13 largest contributors are shown (see text). \textit{Lower panel:} Black crosses represent abundances calculated after replacing only \halflife{} for the 37 nuclei simultaneously. Gray triangles are for the replacement of \pxn{} only.}
\label{fig:figDataImpact}
\end{figure}

\FloatBarrier

Using the reduced set of trajectories, we studied the impact of our data for individual decays. Abundance calculations were performed replacing in the REACLIB database both \halflife{} and \pxn{} for each of the 37 nuclei investigated in this work one at a time. The result (gray triangles) is shown in the upper panel of Fig.~\ref{fig:figDataImpact}, as ratios to the abundance calculated for the base calculation $Y^{R} (Z)$ (original REACLIB, blue squares). For clarity, only the 13 nuclei changing the abundance of any element from Br to Mo by more than 3\% are included. As can be seen, no single beta decay exerts a dominant effect. The global abundance change (represented by the red circles) is the cumulative effect of a number of nuclei depending on the element. It is important to note that the global effect is not merely the addition of individual effects, indicating that correlations between different decays are important. The main contributors to the calculated abundance change for Y, Zr, Nb and Mo are, to a varying degree, \myisoSimp{84}{Zn}, \myisoSimp{85,87}{Ga}, \myisoSimp{86,88}{Ge}, \myisoSimp{89,90,91}{As}, and \myisoSimp{92}{Se}. 

We also studied the relative weight between the impact of new decay rates and new delayed-neutron branchings on the overall abundance change. In view of the previous result, only two calculations were compared. In the first case, \halflife{} for all 37 nuclei was replaced in the REACLIB database with our new data while keeping \pxn{}; in the second case, all \pxn{} were updated and \halflife{} were kept. The result is shown in the lower panel of Fig.~\ref{fig:figDataImpact}. It can be observed that, compared to the replacement of all \halflife{} (black crosses), the replacement of all \pxn{} (gray triangles) has a small effect on abundances, except for Sr. However, they exert a strong influence on shaping the final abundances. For example, by reducing to one-half the impact of half-lives for Y, or cancelling the effect for Sr. This points again to the need for including correlations in sensitivity studies of the impact of decay data in abundance calculations.

\end{document}